\documentclass[12pt, a4paper]{article}
\usepackage{amsmath,amssymb,amsfonts}
\usepackage[colorlinks]{hyperref}
\usepackage{graphicx}

\begin{document}

\title{Non-Euclidean geometry, nontrivial topology and quantum
vacuum effects}

\author{Yurii A. Sitenko$^1$ , Volodymyr M. Gorkavenko$^2$\\
\it \small ${}^{1}$Bogolyubov Institute for Theoretical Physics,
  National Academy of Sciences of Ukraine,\\ \it \small
  14-b Metrologichna str., Kyiv 03143,
 Ukraine\\
 \it \small ${}^{2}$Department of Physics, Taras Shevchenko
National University of Kyiv,\\ \it \small 64 Volodymyrs'ka str.,
Kyiv
 01601, Ukraine}

\date{}
\maketitle

\abstract{Space out of a topological defect of the
Abrikosov-Nielsen-Olesen vortex type is locally flat but
non-Euclidean. If a spinor field is quantized in such a space, then
a variety of quantum effects is induced in the vacuum. On the basis of the
continuum model for long-wavelength electronic excitations,
originating in the tight-binding approximation for the nearest-neighbor 
interaction of atoms in the crystal lattice, we consider
quantum ground-state effects in Dirac materials with two-dimensional
monolayer structures warped into nanocones by a disclination; the nonzero 
size of the disclination is taken into account, and a boundary condition 
at the edge of the disclination is chosen to ensure self-adjointness of 
the Dirac-Weyl Hamiltonian operator. We show that the quantum ground-state 
effects are independent of the disclination size and find circumstances when 
they are independent of parameters of the boundary condition.}

\phantom{hvhv}

\section{Introduction}

Usually, the effects of non-Euclidean geometry are identified with
the effects which are due to the curvature of space. It can be
immediately shown that this is not the case and there are spaces
which are flat but non-Euclidean; moreover, such spaces are of
nontrivial topology.

A simplest example is given by a two-dimensional space (surface)
which is obtained from a plane by cutting a segment of a certain
angular size and then sewing together the edges. The resulting
surface is the conical one which is flat but has a singular
point corresponding to the apex of the cone. To be more precise,
the intrinsic (Gauss) curvature of the conical surface is
proportional to the two-dimensional delta-function placed at the
apex; the coefficient of proportionality is the deficit angle.
Topology of the conical surface with a deleted apex is nontrivial:
$\pi_1=\mathbb{Z}$, where $\pi_1$ is the first homotopy group and
$\mathbb{Z}$ is the set of integer numbers. Usual cones correspond
to positive values of the deficit angle,
i.e. to the situation when a segment is deleted from the plane.
But one can imagine a situation when a segment is added to the
plane; then the deficit angle is negative, and the resulting flat
surface can be denoted as a saddle-like cone. The deleted segment
is bounded by the value of $2\pi$, whereas the added segment is
unbounded. Thus, deficit angles for possible conical surfaces
range from $-\infty$ to $2\pi$.

It is evident that an apex of the conical surface with the
positive deficit angle can play a role of the convex lens, whereas
an apex of the conical surface with the negative deficit angle can
play a role of the concave lens. Really, two parallel trajectories
coming from infinity towards the apex from different sides of it,
after bypassing it, converge (and intersect) in the case of the
positive deficit angle, and diverge in the case of the negative
deficit angle. This demonstrates the non-Euclidean nature of
conical surfaces, providing a basis for understanding such physical
objects as cosmic strings.

Conical space emerges inevitably as an outer space of any
topological defect in the form of the Abrikosov-Nielsen-Olesen
(ANO) vortex \cite{Abr,NO}. Topological defects are produced as a
consequence of phase transitions with spontaneous breakdown of
continuous symmetry in various  physical systems, in particular, in
superfluids, superconductors and liquid crystals. Cosmic strings with a
specific gravitational lensing effect (doubling the image of an
astrophysical object) are the ANO vortices which are produced as a
result of phase transitions at the early stage of evolution of the
Universe, see reviews in Refs.\cite{Hi,Vi}. Otherwise, in micro- and
nanophysics, a wealth of new phenomena, suggesting possible
applications to technology and industry, is promised by a synthesis
in this century of strictly two-dimensional atomic crystals
(for instance, a monolayer of carbon atoms, graphene, \cite{Nov,Ge}).
Topological defects (disclinations) on such layers are similar to
the transverse sections of cosmic strings. A disclination warps a
sheet of a layer, rolling it into a nanocone; moreover, a physically
meaningful range of values of the deficit angle is extended to
include also negative values which correspond to saddle-like cones
or cosmic strings with negative tension.

While considering the effect of the ANO vortex on the vacuum of
quantum matter, the following circumstance should be taken into account:
since the vacuum of quantum matter exists outside the ANO vortex
core, an issue of the choice of boundary conditions at the edge of
the core is of primary  importance. The most general boundary condition
for the matter wave function at the core edge is given by requiring
self-adjointness of the Hamiltonian operator (energy operator in
first-quantized theory).

In the present paper we consider quantum effects which are induced
in the vacuum of the second-quantized pseudorelativistic gapless
(i.e. massless) spinor field in $(2+1)$-dimensional space-time which
is a section orthogonal to the ANO vortex axis; hence the
Hamiltonian operator takes form $H=-i\mbox{\boldmath
$\alpha$}\!\cdot\! \mbox{\boldmath $\nabla$} $, where covariant
derivative $\mbox{\boldmath $\nabla$}$ includes both the affine and
bundle connections (natural units ${\hbar}=v_F=1$ are used, with the
Fermi velocity, $v_F$, becoming the velocity of light, $c$, in the
truly relativistic case). Condensed matter systems with such a behavior 
of low-energy electronic excitations are known as the two-dimensional 
Dirac materials comprising a diverse set ranging from honeycomb crystalline 
structures (graphene \cite{Nov}, silicene and germanene \cite{Cah}) to 
high-temperature $d$-wave superconductors, superfluid phases of helium-3 
and topological insulators, see review in \cite{Weh}. We focus on the quantum 
ground-state effects (induced $R$-current and pseudomagnetic field) of electronic 
excitations in graphitic nanocones, although our consideration is quite 
general to also be relevant for nanocones of the nongraphitic nature as well; 
the finite size of a disclination at the conical apex is taken into account.

\section{Continuum model description of electronic excitations in monolayer atomic crystals with a disclination}

The squared length element of the conical surface is
\begin{equation}\label{1.2}
    ds^2= dr^2+\nu^{-2} r^2 d\varphi^2,
\end{equation}
where $\nu=(1-\eta)^{-1}$ and $2\pi\eta$ is the deficit angle. In
the case of cosmic strings, the present-day astrophysical
observations restrict the values of parameter $\eta $ to range
$0< \eta < 10^{-6}$ (see, e.g., \cite{Bat}). A natural way of 
producing local curvature in the honeycomb lattice of graphene, 
silicene or germanene is by substitung some of hexagons by 
pentagons (positive curvature) and heptagons (negative curvature). 
Thus, in the case of crystalline nanocones, parameter $\eta$ 
takes discrete values: $\eta=N_d/6$,
where $N_d$ is an integer which is smaller than 6. A disclination in
the honeycomb lattice results from a substitution of a
hexagon by a polygon with $6-N_d$ sides; polygons with $N_d>0$
($N_d<0$) induce locally positive (negative) curvature, whereas the
crystalline sheet is flat away from the disclination, as is the
conical surface away from the apex. In the case of nanocones with
$N_d>0$, the value of $N_d$ is related to apex angle $\delta$,
$\sin\frac\delta2=1-\frac{N_d}{6}$, and $N_d$ counts the number of
sectors of the value of $\pi/3$ which are removed from the crystalline
sheet. If $N_d<0$, then $-N_d$ counts the number of such sectors
which are inserted into the crystalline sheet. Certainly, polygonal
defects with $N_d>1$ and $N_d<-1$ are mathematical abstractions,
as are cones with a pointlike apex. In reality, the defects are
smoothed, and $N_d>0$ counts the number of the pentagonal defects
which are tightly clustered producing a conical shape; graphitic
nanocones with the apex angles
$\delta=112.9^\circ,\,83.6^\circ,\,60.0^\circ,\,38.9^\circ,\,19.2^\circ$,
which correspond to the values $N_d=1,\,2,\,3,\,4,\,5$, were
observed experimentally, see \cite{Heiberg} and references therein.
Theory also predicts an infinite series of the saddle-like nanocones
with quantity $-N_d$ counting the number of the heptagonal defects
which are clustered in their central regions. Saddle-like nanocones
serve as an element which is necessary for joining parts of carbon
nanotubes of differing radii. On the basis of the long-wavelength
continuum model originating in the tight-binding approximation for
the nearest-neighbor interactions in the honeycomb crystalline 
lattice, it was proved \cite{SiV7} that the bundle connection 
effectively appears in addition to the affine connection of the 
nanocone, and the Hamiltonian operator takes form
\begin{equation}\label{1.14}
H=-{\rm i}
\left[\alpha^r\left(\partial_r+\frac{1}{2r}\right)+\alpha^\varphi\left(\partial_\varphi-{\rm
i}\frac{{\Phi}}{2\pi}\right)\right],
\end{equation}
where
\begin{equation}\label{1.10a}
\Phi= 3\pi (1- \nu^{-1}) R,
\end{equation}
and matrix $R$ exchanges the sublattice indices, as well as the valley indices, and commutes with $H$ \eqref{1.14}.
Note that in the case of a cosmic string quantity $\Phi$ is the flux of a gauge vector field corresponding to the generator of a spontaneously broken continuous symmetry.
 Both $R$ and $\alpha$-matrices  can be chosen in the block-diagonal form,
\begin{equation}\label{1.15}
R=\left(\begin{array}{cc}
                    I& 0\\
                    0& - I
                    \end{array}\right), \,\,
\alpha^r=\alpha_r=-\left(\begin{array}{cc}
                    \sigma^{2}& 0\\
                   0&\sigma^{2}
                     \end{array}\right),\,\,
\alpha^\varphi=\frac{\nu}r\left(\begin{array}{cc}
                    \sigma^{1}& 0\\
                    0 &-\sigma^{1}
                     \end{array}\right),\,\,
\alpha_\varphi=\frac{r^2}{\nu^2}\,\alpha^\varphi
\end{equation}
($\sigma^{j}$ with $j=1,2,3$ are the Pauli matrices). The solution to the stationary
Dirac-Weyl equation, $ H \psi_{E}({\bf x})=E \psi_{E}({\bf x})$, is decomposed as
\begin{equation}\label{1.16}
\psi_E(\textbf{x}) = \sum_{n \in \mathbb{Z}}
                   \left(\begin{array}{c}
                   f_n^{+}(r,E )e^{ {\rm i} (n+1/2)\varphi} \\
                   g_n^{+}(r,E )e^{ {\rm i} (n+1/2)\varphi} \\
                   f_n^{-}(r,E )e^{ {\rm i} (n-1/2)\varphi} \\
                   g_n^{-}(r,E )e^{ {\rm i} (n-1/2)\varphi}
                    \end{array}\right),
\end{equation}
where the radial function satisfy the system of first-order differential equations
\begin{equation}\label{1.17}
 \left\{
 \begin{array}{c}
\left[-\partial_r +\frac1r (\pm \nu n-\nu+1)\right] f_n^{\pm}(r,E) =E g_n^{\pm}(r,E) \\
\left[\partial_r +\frac1r (\pm \nu n-\nu+2)\right] g_n^{\pm}(r,E)
=E f_n^{\pm}(r,E)
\end{array}
\right\}.\end{equation}

Let us consider nanocones with  $N_d=1,\,2,\,3,\,4,\,5$ $\,\,$
($1<\nu<7$), as well as with  $N_d=-1,\,-2,\,-3$ $\,\,$
($\frac{3}{5} < \nu< 1$), and introduce positive quantity
\begin{equation}\label{1.7}
F=\frac32 \nu - \frac12 \nu {\rm sgn} (\nu - 1) - 1,
\end{equation}
which exceeds $1$ at $N_d=3,\,4,\,5$ $\,\,$ ($2\leq\nu<7$) only;
here $\rm{sgn}(u)$ is the sign function. From the whole variety of
quantum effects in the ground state of electronic excitations (see
\cite{SiV7,SiV1,SiV2}), our focus will be on the induced specific
current ($R$-current) which is defined by expression
\begin{equation}\label{1.7a}
\textbf{j}(\textbf{x})=-\frac12 \int\limits_{-\infty}^\infty dE E
\mbox{$\psi$}^\dag _E(\textbf{x}) \mbox{\boldmath $\alpha$} R  \mbox{$\psi$} _E(\textbf{x})
\end{equation}
The pseudomagnetic field strength, $\textbf{B}_{\rm
I}(\textbf{x})$, is also induced in the ground state, as a consequence of
the analogue of the Maxwell equation,
\begin{equation}\label{1.8}
\mbox{\boldmath $\partial$}\times \textbf{B}_{\rm I}(\textbf{x}) =
e\, \textbf{j}(\textbf{x}).
\end{equation}
Using \eqref{1.15} and \eqref{1.16}, one gets $j_r =0$, and the only component of the induced ground state current,
\begin{equation}\label{1.19}
j_\varphi(r) = - \frac r\nu \int\limits_{-\infty}^\infty dE E\sum_{n \in \mathbb{Z}} [f_n^{+}(r,E) g_n^{+}(r,E) + f_n^{-}(r,E) g_n^{-}(r,E)],
\end{equation}
is independent of the angular variable. The induced ground state
field strength is also independent of the angular variable, being directed orthogonally to the conical surface,
\begin{equation}\label{1.20}
B_{\rm I}(r) = e \nu \int\limits_r^{r_{\rm max}} \frac{dr'}{r'} \,
j_\varphi(r')+ B_{\rm I}(r_{\rm max}).
\end{equation}
with the total flux
\begin{equation}\label{1.21}
\Phi_{\rm I} = \frac{2\pi}\nu \int\limits_{r_0}^{r_{\rm max}} dr\, r B_{\rm
I}(r),
\end{equation}
where it is assumed without a loss of generality that a nanocone is of a rotationally invariant shape with $r_{\rm max}$
being its radius and $r_0$ being the radius of a disclination, $r_{\rm max} \gg r_0$.

\section{Self-adjointness and choice of boundary conditions}

Let us note first, that \eqref{1.14} is not
enough to define the Hamiltonian operator rigorously in a
mathematical sense. To define an operator in a unambiguous way, one
has to specify its domain of definition. Let the set of functions $\psi$
be the domain of definition of operator $H$, and the set of functions
$\tilde \psi$ be the domain of definition of its adjoint, operator
$H^\dag$. Then the operator is Hermitian (or symmetric in
mathematical parlance),
\begin{equation}\label{1.10}
\int\limits_X d^2x \sqrt{g} \,{\tilde\psi}^\dag (H\psi)=
\int\limits_X d^2x \sqrt{g}\,(H^\dag \tilde \psi)^\dag \psi,
\end{equation}
if relation
\begin{equation}\label{1.11}
-{\rm i} \int\limits_{\partial X} d \mbox{\boldmath l}\,
{\tilde \psi}^\dag \mbox{\boldmath $\alpha$} \psi =0
\end{equation}
is valid; here functions $\psi(\textbf{x})$ and $\tilde
\psi(\textbf{x})$ are defined in space $X$ with boundary $\partial
X$. It is evident that condition \eqref{1.11} can be satisfied by
imposing different boundary conditions for $\psi$  and $\tilde
\psi$.  But, a nontrivial task is to find a possibility that a
boundary condition for $\tilde \psi$ is the same as that for $\psi$;
then the domain of definition of $H^\dag$ coincides with that of
$H$, and operator $H$ is self-adjoint (for a review of the Weyl-von
Neumann theory of self-adjoint operators see \cite{Neu,Ree}). The action
of a self-adjoint operator results in functions belonging to its
domain of definition only, and a multiple action and functions of
such an operator,  for instance, the resolvent and evolution operators,
can be consistently defined. Thus, in the case of a surface of radius 
$r_{\rm max}$ with a deleted central disc of radius $r_0$, we have to ensure 
the validity of relations
\begin{equation}\label{3.1}
\left.\tilde\psi^\dag\alpha^r \psi\right|_{r=r_0}=0, \quad \left.\tilde\psi^\dag\alpha^r \psi\right|_{r=r_{\rm max}}=0,
\end{equation}
meaning that the quantum matter excitations do not penetrate outside. It is implied that functions $\psi$ 
and $\tilde \psi$ are differentiable and square-integrable. As $r_{\rm max}\rightarrow \infty$, they conventionally 
turn into differentiable functions corresponding to the continuum, and the condition at $r=r_{\rm max}$ yields
 no restriction at $r_{\rm max}\rightarrow \infty$, whereas the condition at $r=r_0$ yields
\begin{equation}\label{3.2}
\left. \psi\right|_{r=r_0}=\left. K\psi \right|_{r=r_0}, \quad
\left. \tilde\psi\right|_{r=r_0}=\left. K\tilde \psi
\right|_{r=r_0},
\end{equation}
where $K$ is a matrix (element of the Clifford algebra with generators $\alpha^r, \alpha^{\varphi}, \beta$) which obeys
condition
\begin{equation}\label{3.3}
K^2=I
\end{equation}
and   without a loss of generality can be chosen to be Hermitian; in
addition, it has to obey either condition
\begin{equation}\label{3.4}
[K,\alpha^r]_+=0,
\end{equation}
or condition
\begin{equation}\label{3.5}
[K,\alpha^r]_-=0.
\end{equation}
One can simply go through four linearly independent elements of the
Clifford algebra and find that two of them satisfy \eqref{3.4} and
two  other satisfy \eqref{3.5}. However, if one chooses
\begin{equation}\label{3.6}
K=c_1 I+ c_2 \alpha^r
\end{equation}
to satisfy \eqref{3.5}, then \eqref{3.3} is violated. There remains
the only possibility to choose
\begin{equation}\label{3.7}
K= c_1 \beta + c_2 {\rm i} \beta \alpha^r
\end{equation}
with real coefficients obeying condition
\begin{equation}\label{3.8}
c_1^2+c_2^2 = 1;
\end{equation}
then both \eqref{3.3} and \eqref{3.4} are satisfied. Using obvious
parametrization $$ c_1 = \sin\theta,\quad c_2= \cos\theta, $$ we
finally obtain
\begin{equation}\label{3.9}
K = {\rm i} \beta \alpha^r e^{-{\rm i}\theta \alpha^r}.
\end{equation}
Thus, boundary condition \eqref{3.2} with $K$ given by \eqref{3.9}
is the most general boundary condition ensuring self-adjointness
of the Hamiltonian operator on a surface with a deleted disc of radius $r_0$, and parameter $\theta$ can be
interpreted as the self-adjoint extension parameter. Value
$\theta=0$ corresponds to the MIT bag boundary condition which was
proposed as the condition ensuring the confinement of the matter
field, that is, the absence of the matter flux across the boundary
\cite{Joh}. However, it should be comprehended  that a condition
with an arbitrary value of $\theta$ is motivated equally as well as
that with $\theta=0$.

Imposing the boundary condition \eqref{3.2} with matrix $K$
\eqref{3.9} on the solution to the Dirac-Weyl equation,
$\psi_E(\textbf{x})$ \eqref{1.16}, we obtain the condition for the
modes:
\begin{equation}\label{3.10}
\cos\left(\frac{\theta}{2}+\frac{\pi}4 \right)
f_n^\pm(r_0,E)=-\sin\left(\frac{\theta}{2}+\frac{\pi}4
\right)g_n^\pm(r_0,E).
\end{equation}

Let us  compare this with the case of an infinitely thin (pointlike) disclination which was considered in detail in
 \cite{SiV7,SiV1,SiV2}. In the latter case several partial Hamiltonian operators are self-adjoint extended,
  and the deficiency index can be $(0,0)$ (no need for extension, all partial operators are essentially self-adjoint),
   $(1,1)$ (one partial operator is extended with one parameter), $(2,2)$  (two partial operators are extended with four parameters),
    etc. In particular, in the case of carbon nanocones, there is no need for self-adjoint extension for $N_d=3,4,5$, there is one self-adjoint
    extension parameter for $N_d=2,1,-1$, $-2$, $-3,$ $-6$, there are four and more self-adjoint extension parameters for $N_d=-4,-5$ and $N_d\leq -7$.
    For the deficiency index equal to $(1,1)$, the boundary condition at the location of a pointlike disclination $(r=0)$ takes form
\begin{equation}\label{2.11}
    \lim_{r\rightarrow0}
    \left(\frac{r}{r_{\rm max}}\right)^F \cos\left(\frac{\Theta}{2}+\frac{\pi}{4}\right) f_{n_{\rm
    c}}^\pm (r,E)=-\lim_{r\rightarrow 0} \left(\frac{r}{r_{\rm max}}\right)^{1-F}\sin\left(\frac{\Theta}{2}+\frac{\pi}{4}\right) 
    g_{n_{\rm
    c}}^\pm (r,E),
\end{equation}
where $\Theta$ is the self-adjoint extension parameter, $F$ is given
by \eqref{1.7} for $N_d=2,1,-1$, $-2$, $-3$ and $F=1/2$ for
$N_d=-6$,
 while $n_c=\pm\frac12[{\rm sgn}(\nu-1)-1]$ for $N_d=2,1,-1$, $-2$, $-3$ and $n_c=\mp2$ for $N_d=-6$. As follows
from the present section, in the case of a disclination of nonzero size, when the boundary condition is imposed at its edge,
the total Hamiltonian operator is self-adjoint extended with the use of one parameter, see \eqref{3.10}.

\section{Quantum effects in the ground state of electronic excitations in nanocones}

Using the explicit form of modes $f_{n}^\pm$ and $g_{n}^\pm$,
satisfying \eqref{1.17} and \eqref{3.10}, we calculate current
\eqref{1.19} and field strength \eqref{1.20}. In the case of
$\frac35 < \nu < 2$ $\,\,$ ($0<F<1$) we obtain
\begin{multline}\label{d1}
\left.j_\varphi(r)\right|_{F<\frac12,\theta\neq
-\frac\pi2}=-\frac{1}{(2\pi)^2}\frac1r
\left\{
\int\limits_0^\infty \frac{du}{\cosh^2(u/2)} \right.\\  \times
\frac{\sin(F\pi)
\cosh\left[\left(F+\nu-\frac12 \right)u\right]- \sin[(F+\nu)\pi)]
\cosh\left[\left(F-\frac12\right)u\right]}{\cosh(\nu u)-\cos(\nu
\pi)}\\
+8r^2\int\limits_0^\infty dq\,q\left[\sum_{l=0}^\infty
C^{(\wedge)}_{\nu l+1-F}(qr_0)K_{\nu l+1-F}(qr)
K_{\nu l-F}(qr)\right.\\
-\left.\left.\sum_{l=1}^\infty
C^{(\vee)}_{\nu l+F}(qr_0)K_{\nu l+F}(qr)
K_{\nu l-1+F}(qr)\right]\right\},
\end{multline}\vspace{-1em}
\begin{multline}\label{d2}
\left.j_\varphi(r)\right|_{F>\frac12,\theta\neq
\frac\pi2}=\frac{1}{(2\pi)^2}\frac1r
\left\{
\int\limits_0^\infty \frac{du}{\cosh^2(u/2)} \right.\\  \times
\frac{\sin(F\pi)
\cosh\left[\left(F-\nu-\frac12 \right)u\right]- \sin[(F-\nu)\pi)]
\cosh\left[\left(F-\frac12\right)u\right]}{\cosh(\nu u)-\cos(\nu
\pi)}\\
-8r^2\int\limits_0^\infty dq\,q\left[\sum_{l=1}^\infty
C^{(\wedge)}_{\nu l+1-F}(qr_0)K_{\nu l+1-F}(qr)
K_{\nu l-F}(qr)\right.\\
-\left.\left.\sum_{l=0}^\infty
C^{(\vee)}_{\nu l+F}(qr_0)K_{\nu l+F}(qr)
K_{\nu l-1+F}(qr)\right]\right\},
\end{multline}\vspace{-1em}
\begin{multline}\label{d3}
\left.j_\varphi(r)\right|_{F\neq\frac12,\theta=\pm
\frac\pi2}=\mp\frac{1}{2(2\pi)^2}\frac1r
\left\{
\int\limits_0^\infty \frac{du}{\cosh^2(u/2)} \right.\\  \times
\frac{\sin(F\pi) \cosh\left[\left(F-\frac12\pm
\nu\right)u\right]- \sin\left[\left(F \pm
\nu\right)\pi\right]\cosh\left[\left(F-\frac12\right)u\right]}{\cosh(\nu u)-\cos(\nu
\pi)}\\
+8r^2\int\limits_0^\infty
dq\,q\left[\frac{I_{\frac12\mp\left(F-\frac12\right)}(qr_0)}{K_{\frac12\mp\left(F-\frac12\right)}(qr_0)}
K_{F}(qr)
K_{1-F}(qr)\right.\\
+\sum_{l=1}^\infty\Biggl(
\frac{I_{\nu l-F+\frac12\pm\frac12}(qr_0)}{K_{\nu l-F+\frac12\pm\frac12}(qr_0)}
K_{\nu l+1-F}(qr)
K_{\nu l-F}(qr)\Biggr.\\
\left.\left.\Biggl.+\frac{I_{\nu l+F-\frac12\mp\frac12}(qr_0)}{K_{\nu l+F-\frac12\mp\frac12}(qr_0)}
K_{\nu l+F}(qr)
K_{\nu l-1+F}(qr)\Biggr)\right]\right\},
\end{multline}\vspace{-1em}
\begin{equation}\label{d4}
\left.j_\varphi(r)\right|_{F=\frac12}=-\frac{\sin\theta}{2\pi^2}\left[\frac1{r-r_0}+8r\int\limits_0^\infty
dq\,q \sum_{l=1}^\infty \tilde C_{\nu
l+\frac12}(qr_0) K_{\nu l+\frac12}(qr)K_{\nu
l-\frac12}(qr)\right],
\end{equation}\vspace{-1em}
\begin{multline}\label{d5a}
\left.B_{\rm I}(r)\right|_{F<\frac12,\theta\neq
-\frac\pi2}=-\frac{\nu e}{(2\pi)^2}\frac1r
\left\{
\int\limits_0^\infty \frac{du}{\cosh^2(u/2)} \right.\\  \times
\frac{\sin(F\pi)
\cosh\left[\left(F+\nu-\frac12 \right)u\right]- \sin[(F+\nu)\pi)]
\cosh\left[\left(F-\frac12\right)u\right]}{\cosh(\nu u)-\cos(\nu
\pi)}\\
+8r\int\limits_r^{r_{max}} dr'\int\limits_0^\infty
dq\,q\left[\sum_{l=0}^\infty
C^{(\wedge)}_{\nu l+1-F}(qr_0)K_{\nu l+1-F}(qr')
K_{\nu l-F}(qr')\right.\\
-\left.\left.\sum_{l=1}^\infty
C^{(\vee)}_{\nu l+F}(qr_0)K_{\nu l+F}(qr')
K_{\nu l-1+F}(qr')\right]\right\},
\end{multline}\vspace{-1em}
\begin{multline}\label{d6}
\left.B_{\rm I}(r)\right|_{F>\frac12,\theta\neq \frac\pi2}=\frac{\nu
e}{(2\pi)^2}\frac1r \left\{
\int\limits_0^\infty \frac{du}{\cosh^2(u/2)} \right.\\  \times
\frac{\sin(F\pi)
\cosh\left[\left(F-\nu-\frac12 \right)u\right]- \sin[(F-\nu)\pi)]
\cosh\left[\left(F-\frac12\right)u\right]}{\cosh(\nu u)-\cos(\nu
\pi)}\\
-8r\int\limits_{r}^{r_{max}} dr'\int\limits_0^\infty
dq\,q\left[\sum_{l=1}^\infty
C^{(\wedge)}_{\nu l+1-F}(qr_0)K_{\nu l+1-F}(qr')
K_{\nu l-F}(qr')\right.\\
-\left.\left.\sum_{l=0}^\infty
C^{(\vee)}_{\nu l+F}(qr_0)K_{\nu l+F}(qr')
K_{\nu l-1+F}(qr')\right]\right\},
\end{multline}\vspace{-1em}
\begin{multline}\label{d7}
\left.B_{\rm I}(r)\right|_{F\neq\frac12,\theta=\pm
\frac\pi2}=\mp\frac{\nu e}{(2\pi)^2}\frac1r
\left\{
\int\limits_0^\infty \frac{du}{\cosh^2(u/2)} \right.\\  \times
\frac{\sin(F\pi) \cosh\left[\left(F-\frac12\pm
\nu\right)u\right]- \sin\left[\left(F \pm
\nu\right)\pi\right]\cosh\left[\left(F-\frac12\right)u\right]}{\cosh(\nu u)-\cos(\nu
\pi)}\\
+8r\int\limits_{r}^{r_{max}} dr'\int\limits_0^\infty
dq\,q\left[\frac{I_{\frac12\mp\left(F-\frac12\right)}(qr_0)}{K_{\frac12\mp\left(F-\frac12\right)}(qr_0)}
K_{F}(qr')
K_{1-F}(qr')\right.\\
+\sum_{l=1}^\infty\Biggl(
\frac{I_{\nu l-F+\frac12\pm\frac12}(qr_0)}{K_{\nu l-F+\frac12\pm\frac12}(qr_0)}
K_{\nu l+1-F}(qr')
K_{\nu l-F}(qr')\Biggr.\\
\left.\Biggl.\left.+\frac{I_{\nu l+F-\frac12\mp\frac12}(qr_0)}{K_{\nu l+F-\frac12\mp\frac12}(qr_0)}
K_{\nu l+F}(qr')
K_{\nu l-1+F}(qr')\Biggr)\right]\right\}
\end{multline}\vspace{-1em}
and
\begin{multline}\label{d8}
\left.B_{\rm I}(r)\right|_{F=\frac12}=\frac{\nu e
\sin\theta}{2\pi^2}\!\left[\!\frac1{r_0}\ln\left(1\!-\!\frac{r_0}r\right)\right.
\\ \left.-8\!\int\limits_{r}^{r_{max}}\! dr'\! \int\limits_0^\infty\! dq\,q
\sum_{l=1}^\infty \tilde C_{\nu
l+\frac12}(qr_0)\! K_{\nu l+\frac12}(qr')K_{\nu
l-\frac12}(qr')\right]\!,
\end{multline}\vspace{-1em}
where
\begin{multline}\label{4.6}
C^{(\wedge)}_\rho(v)=\left\{I_\rho(v)K_\rho(v)\tan\left(\frac\theta2+\frac\pi4\right)
-I_{\rho-1}(v)K_{\rho-1}(v)\cot\left(\frac\theta2+\frac\pi4\right)
\right\} \\
\times\left[K^2_\rho(v)\tan\left(\frac\theta2+\frac\pi4\right) + K^2_{\rho-1}(v)\cot\left(\frac\theta2+\frac\pi4\right)
\right]^{-1},
\end{multline}
\begin{multline}\label{4.7}
C^{(\vee)}_\rho(v)=\left\{I_\rho(v)K_\rho(v)\cot\left(\frac\theta2+\frac\pi4\right)-
I_{\rho-1}(v)K_{\rho-1}(v)\tan\left(\frac\theta2+\frac\pi4\right)
\right\} \\
\times\left[K^2_\rho(v)\cot\left(\frac\theta2+\frac\pi4\right) + K^2_{\rho-1}(v)\tan\left(\frac\theta2+\frac\pi4\right)
\right]^{-1}
\end{multline}
and
\begin{equation}\label{4.21}
{\tilde C}_{\nu l+\frac12}(v)=\frac2v\frac{K_{\nu
l+\frac12}(v)K_{\nu l-\frac12}(v)}{\cos^2\theta\left[K_{\nu l+\frac12}^2(v)+K_{\nu l-\frac12}^2(v)\right]^2+4\sin^2\theta \,K^2_{\nu l+\frac12}(v)K^2_{\nu
l-\frac12}(v)};
\end{equation}
$I_\rho(u)$ and $K_\rho(u)$ are the modified Bessel functions with
the exponential increase and decrease, respectively, at large real
positive values of their argument.

In the case of $2\leq\nu<7$ $\,\,$ ($F=\nu-1$) we obtain
\begin{multline}\label{d5}
j_\varphi(r)=-\frac{1}{(2\pi)^2}\frac1r
\left\{\frac{2\pi}{\nu}\sum_{p=1}^{\left[\!\left| {\nu}/2
\right|\!\right]} \frac{\sin(3p\pi/\nu)}{\sin^2(p\pi/\nu)}-\frac\pi\nu \delta_{\nu, \, 2N}
+\sin(\nu\pi)\int\limits_0^\infty \frac{du}{\cosh^2(u/2)}\right.\\
\times\frac{
\cosh\left(\frac32 u \right)}{\cosh(\nu u)-\cos(\nu
\pi)}
+8r^2\int\limits_0^\infty dq\,q\left[\sum_{l=1}^\infty
C^{(\wedge)}_{\nu (l-1)+2}(qr_0)K_{\nu (l-1)+2}(qr)
K_{\nu (l-1)+1}(qr)\right.\\
-\left.\left.\sum_{l=0}^\infty
C^{(\vee)}_{\nu (l+1)-1}(qr_0)K_{\nu (l+1)-1}(qr)
K_{\nu (l+1)-2}(qr)\right]\right\}
\end{multline}\vspace{-1em}
and
\begin{multline}\label{d6a}
B_{\rm I}(r)=-\frac{\nu e}{(2\pi)^2}\frac1r
\left\{\frac{2\pi}{\nu}\sum_{p=1}^{\left[\!\left| {\nu}/2
\right|\!\right]} \frac{\sin(3p\pi/\nu)}{\sin^2(p\pi/\nu)}-\frac\pi\nu \delta_{\nu, \, 2N}
+\sin(\nu\pi)\int\limits_0^\infty \frac{du}{\cosh^2(u/2)}\right.\\
\times\frac{
\cosh\left(\frac32 u \right)}{\cosh(\nu u)-\cos(\nu
\pi)}
+8r\int\limits_{r}^{r_{max}} dr'\int\limits_0^\infty
dq\,q\left[\sum_{l=1}^\infty
C^{(\wedge)}_{\nu (l-1)+2}(qr_0)K_{\nu (l-1)+2}(qr')
K_{\nu (l-1)+1}(qr')\right.\\
-\left.\left.\sum_{l=0}^\infty
C^{(\vee)}_{\nu (l+1)-1}(qr_0)K_{\nu (l+1)-1}(qr')
K_{\nu (l+1)-2}(qr')\right]\right\},
\end{multline}
where $\left[\!\left| u \right|\!\right]$ is the integer part of
quantity $u$ (i.e. the integer which is less than or equal to $u$),
$p$ and  $N$ denote positive integers, $\delta_{\omega, \, \omega'}$ is the Kronecker symbol 
($\delta_{\omega, \, \omega'}=0$ at $\omega' \neq \omega$ and 
$\delta_{\omega, \, \omega} = 1$).

It should be noted that the integral over the $q$ variable in \eqref{d1} -- \eqref{d4} and  \eqref{d5} 
vanishes in the limit of $r_0 \rightarrow 0$. Moreover, in the limit of $r \rightarrow \infty$, it decreases 
at least as $r^{-2\rho}$, where
\begin{equation}\label{39}
\rho=2-F, \quad \frac 35 < \nu <2, \quad \left\{\begin{array}{l}
 0<F<\frac12, \quad \theta\neq -\frac\pi2 \\
 \frac12<F <1, \quad \theta =\frac\pi2
\end{array}\right\},
\end{equation}
\begin{equation}\label{40}
\rho=1+F, \quad \frac 35 < \nu <2, \quad \left\{\begin{array}{l}
 \frac12<F<1, \quad \theta\neq \frac\pi2 \\
 0<F <\frac12, \quad \theta =-\frac\pi2
\end{array}\right\},
\end{equation}
\begin{equation}\label{41}
\rho=\nu+\frac 12, \quad \frac 35 < \nu <2, \quad F=\frac12,
\end{equation}
\begin{equation}\label{42}
\left\{\begin{array}{l}
 \rho=1+\frac{\ln \ln r}{2\ln r}, \quad \nu=2 \\
 \rho=\nu-1, \quad 2<\nu<7
\end{array}\right\},\quad F=\nu-1.
\end{equation}
The latter circumstance has far-reaching consequences, when we turn to the total flux of the induced ground 
state field strength, see \eqref{1.21}. Namely, the contribution of the $q$-integral to $\Phi_{\rm I}$ is 
damped and the field strength is proportional  to  
the current in the physically sensible case, i.e. at $r_{\rm max} \gg r_0$:
\begin{equation}\label{last1}
j_\varphi(r)=\frac{\Phi_{\rm I}}{2\pi e r_{\rm max}} \,\, \frac1r, \quad 
B_{\rm I}(r)=\frac{\nu\, \Phi_{\rm I}}{2\pi r_{\rm max}} \,\, \frac1r,
\end{equation}
where
\begin{multline}\label{d10}
\left.\Phi_{\rm I} \right|_{0<F<\frac12, \,\, \theta\neq
-\frac\pi2}=\left.\Phi_{\rm I}\right|_{\frac12<F<1, \,\, \theta=\frac\pi2}=
-\frac{e}{2\pi}\,r_{\rm max} \int\limits_0^\infty \frac{du}{\cosh^2(u/2)} \\  \times
\frac{\sin(F\pi)
\cosh\left[\left(F+\nu-\frac12 \right)u\right]- \sin[(F+\nu)\pi)]
\cosh\left[\left(F-\frac12\right)u\right]}{\cosh(\nu u)-\cos(\nu
\pi)}, \quad \frac 35 < \nu <2,
\end{multline}
\begin{multline}\label{d11}
\left.\Phi_{\rm I}\right|_{\frac12<F<1, \,\, \theta\neq
\frac\pi2}=\left.\Phi_{\rm I} \right|_{0<F<\frac12, \,\, \theta=-\frac\pi2}=
\frac{e}{2\pi}\,r_{\rm max} \int\limits_0^\infty \frac{du}{\cosh^2(u/2)} \\  \times
\frac{\sin(F\pi)
\cosh\left[\left(F-\nu-\frac12 \right)u\right]- \sin[(F-\nu)\pi)]
\cosh\left[\left(F-\frac12\right)u\right]}{\cosh(\nu u)-\cos(\nu
\pi)}, \quad \frac 35 < \nu <2,
\end{multline}
\begin{equation}\label{d13}
\left.\Phi_{\rm I} \right|_{F=\frac12}=-\frac{e
\sin\theta}{\pi}r_{\rm max}
\end{equation}
and
\begin{multline}\label{d13}
\left.\Phi_{\rm I} \right|_{F=\nu-1}=-\frac{e}{2\pi}r_{\rm max}
\left[\frac{2\pi}{\nu}\sum_{p=1}^{\left[\!\left| {\nu}/2
\right|\!\right]} \frac{\sin(3p\pi/\nu)}{\sin^2(p\pi/\nu)}-\frac\pi\nu \delta_{\nu, \, 2N} \right. \\
\left. + \sin(\nu\pi)\int\limits_0^\infty \frac{du}{\cosh^2(u/2)}
\frac{\cosh\left(\frac32 u \right)}{\cosh(\nu u)-\cos(\nu
\pi)}\right], \quad 2\leq\nu<7.
\end{multline}

\section{Conclusions}

Quantum vacuum effects which are due to non-Euclidean geometry of nanocones are studied in the present paper. On the basis of the 
continuum model for long-wavelength electronic excitations, originating in the tight-binding approximation for the nearest-neighbor 
interactions of the lattice atoms, we consider quantum ground-state effects in monolayers warped into nanocones by a disclination; 
the nonzero size of the disclination at the apex of a nanocone is taken into account. We show that the $R$-current circulating 
around the disclination is induced in the ground state, see  \eqref{d1} -- \eqref{d4} and  \eqref{d5}. The pseudomagnetic field 
strength which is orthogonal to the nanocone surface is induced in the ground state as well, see \eqref{d5a} -- \eqref{d8} 
and \eqref{d6a}. Both the current and the field strength are invariant under time reversal and consist of two parts: one is 
independent of the disclination size, $r_0$, and another one depending on $r_0$ is damped at large distances from the disclination. 
In the physically sensible case, that is, at $r_{\rm max} \gg r_0$, the latter part is negligible, and we arrive at the conclusion 
that quantum ground-state effects are independent of $r_0$. Moreover, in this case the field strength is proportional to the current, 
see \eqref{last1}, with $\Phi_{\rm I}$ being the total flux through a nanocone with radial size $r_{\rm max}$, 
see \eqref{d10} -- \eqref{d13}.
  
 Our results are relevant for the two-dimensional Dirac materials of conical shape. In particular, for the case of the carbon 
 monolayer (graphene) warped into a nanocone by a disclination, that is, a $(6-N_d)$-gonal $(N_d\neq 0)$ defect inserted in the otherwise 
 perfect two-dimensional hexagonal lattice, the results can be summarized as follows. The dominating contribution to the
 induced ground-state flux of the pseudomagnetic field through carbon nanocones with $N_d=\pm1,\pm2,\pm3,4,5,-6$ is
\begin{equation}\label{last2} 
\left.\Phi_{\rm I} \right|_{\theta\neq -\frac\pi2}\!=\!
-\frac{e}{2\pi}\,r_{\rm max}\!\! \int\limits_0^\infty\!\!
\frac{du}{\cosh^2(u/2)} \, \frac{\sin\left(\frac15\pi\right)
\cosh\left(\frac9{10} u\right)\!-\! \sin\left(\frac75 \pi\right)
\cosh\left(\frac3{10}u\right)}{\cosh\left(\frac65
u\right)-\cos\left(\frac65 \pi\right)},\,\, N_d=1,
\end{equation}
\begin{equation}\label{last3}
\left.\Phi_{\rm I} \right|_{\theta= -\frac\pi2}=
\frac{e}{2\pi}\,r_{\rm max} \sin\left(\frac15\pi\right)
\int\limits_0^\infty \frac{du}{\cosh^2(u/2)}\,
\frac{\cosh\left(\frac3{2} u\right)}{\cosh\left(\frac65
u\right)-\cos\left(\frac65 \pi\right)},\,\, N_d=1,
\end{equation}
\begin{equation}\label{last4}
\left.\Phi_{\rm I} \right|_{\theta\neq \frac\pi2}\!=\!
\frac{e}{2\pi}\,r_{\rm max}\!\! \int\limits_0^\infty\!\!
\frac{du}{\cosh^2(u/2)} \, \frac{\sin\left(\frac57 \pi\right)
\cosh\left(\frac9{14} u\right)\!+\! \sin\left(\frac17 \pi\right)
\cosh\left(\frac3{14}u\right)}{\cosh\left(\frac67
u\right)-\cos\left(\frac67 \pi\right)},\,\, N_d=-1,
\end{equation}
\begin{equation}\label{last5}
\left.\Phi_{\rm I} \right|_{\theta= \frac\pi2}\!=\!
-\frac{e}{2\pi}\,r_{\rm max}\!\! \int\limits_0^\infty\!\!
\frac{du}{\cosh^2(u/2)} \frac{\sin\left(\frac57\pi\right)
\cosh\left(\frac{15}{14} u\right)\!-\! \sin\left(\frac{11}7
\pi\right) \cosh\left(\frac3{14}u\right)}{\cosh\left(\frac67
u\right)-\cos\left(\frac67 \pi\right)},\,\, N_d=-1,
\end{equation}
\begin{equation}\label{last6}
\Phi_{\rm I} =-\frac{e \, \sin\theta}{\pi} \,  r_{\rm max}, \,\, N_d=\pm 2,-6,
\end{equation}
\begin{equation}\label{last7}
\left.\Phi_{\rm I} \right|_{\theta\neq -\frac\pi2}= -\frac{e \,
\sqrt{3}}{4\pi}\,r_{\rm max} \int\limits_0^\infty \frac{du}{\cosh(u/2)}
\frac{1}{\cosh\left(\frac23 u\right)-\cos\left(\frac23
\pi\right)},\,\, N_d=-3,
\end{equation}
\begin{equation}\label{last8}
\left.\Phi_{\rm I} \right|_{\theta= -\frac\pi2}= \frac{e\,
\sqrt{3}}{4\pi}\,r_{\rm max} \int\limits_0^\infty
\frac{du}{\cosh^2(u/2)} \frac{ \cosh\left(\frac{5}{6} u\right)+
\cosh\left(\frac1{6}u\right)}{\cosh\left(\frac23
u\right)-\cos\left(\frac23 \pi\right)},\,\, N_d=-3,
\end{equation}
\begin{align}
&    \Phi_{\rm I}  =\frac{e}{4} \, r_{\rm max}, \,\, N_d=3,\\
&    \Phi_{\rm I}  =0, \,\, N_d=4,\\
&    \Phi_{\rm I}  = -\frac{7 e}{12} \, r_{\rm max},\,\,  N_d=5.
\end{align}

We conclude that the quantum ground-state effects change drastically as $N_d$ changes. The effects are absent in the case of 
the four-heptagonal defect ($N_d= 4$), whereas they appear of opposite signs as a heptagon is removed from ($N_d= 3$)
or added to ($N_d= 5$) this defect, see (55) -- (57). These cases are independent of the boundary parameter, $\theta$; note that namely these cases 
correspond to that situation with the zero-size defect when there is no need for self-adjoint extension (the deficiency index is 
(0,0)). In all other cases the results depend on $\theta$. The most distinct dependence is characteristic for the cases of 
two-pentagonal, two- and six-heptagonal defects, when the results coincide, see \eqref{last6}; note that the electric charge is 
not induced in these cases \cite{SiV7}. In the cases of one-pentagonal, one- and three-heptagonal defects, the results are almost 
independent of $\theta$ unless $\theta=-\frac{\pi}{2}$ for $N_d=1, -3$ and $\theta=\frac{\pi}{2}$ for $N_d=-1$, see 
\eqref{last2} -- \eqref{last5}, \eqref{last7} and \eqref{last8}.
 
Freely suspended samples of crystalline monolayers are not exactly plane surfaces, but possesss ripples which produce pseudomagnetic 
fields causing strains and scattering of electronic excitations in a sample \cite{Voz}. As follows from the present 
consideration, pseudomagnetic fields can be induced in the locally flat regions out of disclinations, and this may have
observable consequences in experimental measurements, likely with the use of scanning tunnel and transmission electron 
microscopy.

\phantom{jvhvj}

{\it Acknowledgements}
Yu.A.S would like to thank the Organizers of the 10th Bolyai-Gauss-Lobachevsky Conference on Non-Euclidean Geometry and its 
Applications for kind hospitality during this interesting and inspiring meeting.






\end{document}